\documentclass[11pt]{article}

\usepackage{ulem}
\usepackage{natbib}
\usepackage{graphicx,amsfonts}
\renewcommand{\Im}{\mathrm{Im}}
\renewcommand{\Re}{\mathrm{Re}}
\newcommand{\etal}{{\it \etal}}

\begin{document}
\title{Thermal fracture as a framework for quasi-static crack propagation}
\author{F. Corson $^\dag$, M. Adda-Bedia$^\dag$, H. Henry$^\ddag$ and E. Katzav$^\dag$\\
$^\dag$Laboratoire de Physique Statistique, ENS, Paris VI, Paris VII, CNRS,\\
24 rue Lhomond, 75005 Paris, France.\\
$^\ddag$Laboratoire de Physique de la Mati\`ere condens\'ee, CNRS UMR 7643,\\
Ecole Polytechnique,  91128 Palaiseau Cedex, France.}
\date{\today}

\maketitle

\begin{abstract}

We address analytically and numerically the problem of crack path prediction in the model system of a crack propagating under thermal loading. We show that one can explain the instability from a straight to a wavy crack propagation by using only the principle of local symmetry and the Griffith criterion. We then argue that the calculations of the stress intensity factors can be combined with the standard crack propagation criteria to obtain the evolution equation for the crack tip within any loading configuration. The theoretical results of the thermal crack problem agree with the numerical simulations we performed using a phase field model. Moreover, it turns out that the phase-field model allows to clarify the nature of the transition between straight and oscillatory cracks which is shown to be supercritical.

\end{abstract}

\section{Introduction}

Crack path prediction is one of the main challenges in the field of fracture mechanics. The reason is simple - a satisfactory equation of motion of a crack tip is associated with a fundamental understanding of material separation mechanisms~\citep{Freund,Broberg,marderreview,adda99,leblond-livre}. From a more general point of view and alongside its importance in many applications, the determination of crack propagation laws is necessary to describe the fracture phenomenon as a pattern formation process induced by mechanical stresses. Within the framework of Linear Elastic Fracture Mechanics (LEFM), the propagation of a crack is mainly governed by the singular behavior of the stress field in the vicinity of its tip~\citep{Freund,Broberg,leblond-livre}. For a two-dimensional quasi-static crack, which is the main purpose of the present work, this behavior is given by
\begin{equation}
\sigma_{ij}(r,\phi)=\frac{K_\mathrm{I}}{\sqrt{2\pi r}}\Sigma_{ij}^\mathrm{I}(\phi)
+\frac{K_\mathrm{II}}{\sqrt{2\pi r}}\Sigma_{ij}^\mathrm{II}(\phi)+O(r^0)\;,
 \label{SIF}
\end{equation}
where $\Sigma_{ij}^\mathrm{I}(\phi)$ and $\Sigma_{ij}^\mathrm{II}(\phi)$ are universal functions describing the angular variation of the stress field, and $K_\mathrm{I}$ and $K_\mathrm{II}$ are the stress intensity factors (SIFs). The evolution of the crack tip is governed by the Griffith energy criterion \citep{Griffith,Freund,Broberg,leblond-livre}, which states that the intensity of the loading necessary to induce propagation is given by $G = \Gamma$, where $G$ is the energy release rate and $\Gamma$ is the fracture energy of the material, that is the energy needed to create new free surfaces. This criterion can be rewritten using the stress intensity factor, in which case it is referred to as the Irwin criterion \citep{Irwin}
\begin{equation}
K_\mathrm{I}=K_\mathrm{Ic}\equiv\sqrt{2\mu\Gamma}\;,
 \label{Irwin}
\end{equation}
where $K_\mathrm{Ic}$ is the toughness of the material and $\mu$ is the shear modulus. The Griffith criterion was originally formulated for quasi-static crack propagation~\citep{Griffith}, and later generalized to rapidly moving cracks \citep{Freund}. While this criterion is very useful in predicting crack initiation, it cannot predict the direction of the crack tip, and therefore in most cases it is not sufficient to determine the actual path of the crack. In order to achieve this, several suggestions have been made. Among them, the Principle of Local Symmetry (PLS) states that the crack advances in such a way that in-plane shear stress vanishes in the vicinity of the crack tip, or explicitly
\begin{equation}
K_\mathrm{II}=0\;,
 \label{eq:PLS0}
\end{equation}
This rule was proposed for in-plane quasi-static cracks \citep{goldstein-74,leblond-89}, and then generalized to rapidly moving cracks \citep{adda99}. Note that from a historical point of view, the first criterion for crack path selection, based on symmetry arguments similar to the PLS, was first formulated for antiplane (mode~III) crack propagation \citep{Barenblatt}. Recently, the dynamics of a rough crack in two dimensions has been successfully described by an equation of motion derived using the PLS \citep{katzav07a}. Another suggestion, based on symmetry arguments, and which recovers the principle of local symmetry in a certain limit, was proposed by \citet{hodgdon-93}, who formulated an equation of motion of the crack tip. However, this formulation introduces additional length scales, which are not known {\em a priori}. A different approach, known as the maximum energy release rate criterion~\citep{Erdogan63} states that the crack advances in a direction that maximizes its energy release rate. Interestingly, even though this approach is quite different from the PLS, it produces very similar results to those obtained using the PLS to such an extent that they were even conjectured to coincide \citep{bilby-75}. However, as shown in \citep{amestoy-92} for kinked cracks and in \citep{katzav07b} for branched cracks, the results are not exactly the same and a clear distinction can be made between the two. Actually, it has been shown that the PLS is the only self-consistent criterion \citep{leblond-89,leblond-livre}. A theoretical challenge is to explain the current rich crack path phenomenology, including the various known instabilities, using a pure LEFM approach combined with the Griffith criterion and the PLS. This has been recently argued for roughening instabilities of cracks in \citep{katzav07a}, as well as for the branching instability in \citep{katzav07b} and here we wish to contribute further to this effort.

Trying to address the problem of crack path prediction from the experimental side, the thermal crack problem has attracted a lot of attention. The propagation of cracks induced by thermal gradients has been widely studied \citep{RonsinPRL,Ronsinthese,yuse-97,ronsin-98,yang-01, Deegan03, yoneyama-06, Sakaue-08, yoneyama-08} since the work of \citet{yuse-93}. In a typical experiment, a glass strip with a notch at its end is pulled at a constant velocity from an oven into a cold bath (see Fig.~\ref{diagram}). The control parameters in this experiment are mainly the width of the strip, the temperature gradient between the oven and the cold bath, and the pulling velocity. If the velocity is small enough, a crack does not propagate. Above a first critical velocity, the crack starts propagating following a straight centered path, and above a second critical velocity, the crack begins to oscillate with a well defined wavelength. However, the nature of the transition from straight to wavy path is not fully understood. For example, the experimental results \citep{Ronsinthese,yuse-97} are not sufficient to determine whether the bifurcation is super-critical or sub-critical (also referred to as continuous or discontinuous transition)\footnote{In~\citep{yuse-97}, the bifurcation is claimed to be supercritical. Nonetheless, data points close to threshold are not accurate enough to rule out any other hypothesis (see Fig. 5 on page 372 in \citep{yuse-97}). Results presented in~\citep{Ronsinthese} suffer from the same lack of data points close to threshold.}. At higher pulling velocities, irregular oscillations and branching are observed. Interestingly, these are very reproducible regimes, which makes the thermal crack an ideal model experiment as it allows to study the slow propagation of cracks under well controlled conditions.

In an attempt to provide a theoretical explanation, \citet{marder-94} successfully determined the propagation threshold by calculating $K_\mathrm{I}$ for a straight crack propagating at the center of the plate. He also proposed a description of the oscillatory instability using the stability criterion derived by \citet{cotterell-80} for a crack in an infinite plate (the celebrated $T$-criterion), but that implied the improbable requirement that the fracture energy should depend strongly on the velocity. Eventually, this result was found to be incompatible with the experimental evidence \citep{RonsinPRL}, and serves as a classical example for the violation of the $T$-criterion. \citet{sasa-94} attempted to predict the threshold of the oscillatory instability and its wavelength by using the PLS under the infinite-plate approximation. Although they gave a reasonable qualitative description, their results deviated systematically from the experimental measurements. The next step was taken by \citet{adda-95}, who calculated $K_\mathrm{II}$ directly for a sinusoidal perturbation in a finite strip, in the configuration where the crack tip is at the center of the strip. However, in order to predict the oscillatory instability they resorted to an additional criterion, namely that the stability with respect to this perturbation depends on the sign of $K_\mathrm{II}$. More recently, \citet{bouchbinder-03} wrote an amplitude equation for a sinusoidal perturbation using the propagation criterion of \citet{hodgdon-93}. One should note that all these approaches rely on additional criteria when determining the instability threshold and cannot do so using only  the classical principles of crack propagation. In addition, theoretical results, except an attempt in \citep{bouchbinder-03}, are mostly limited to the study of the linear stability of a straight crack or to the study of transient regimes. This situation calls for a reexamination of the thermal crack problem in order to explain it using solid arguments.

In this work, we begin with the study of quasistatic cracks in a quenched glass plate using a phase-field approach. Phase field models were originally introduced by~\citet{caginalp86,levine86} to describe the propagation of solidification fronts. Extensions of this work allowed quantitative modeling of dendritic growth~\citep{rappel98} and since then, phase field methods have been applied to many solidification problems such as growth of binary alloys~\citep{plapp} and formation of polycristalline solids~\citep{warren}. Furthermore, they have been extended to other free boundary problems including viscous fingering~\citep{folch} and crack propagation~\citep{Karmafrac,Karmabranching,henrylevine2004,hakim05,hakim08}.
In the latter case,  the phase field approach  turns out to be very similar to the variational approach of fracture  which uses the description of free discontinuity problems  \citep{braides-98,acerbi-99} using the  $\Gamma$-convergence \citep{ambrosio-90,ambrosio-92}. The numerical results of the phase-field model we derive allow to clarify the nature of the transition between straight and oscillatory cracks.

We then provide a theoretical analysis of the experiment of~\citet{yuse-93}. As in previous works \citep{adda-95,bouchbinder-03}, the analysis is based on the calculation of the SIFs. However, we calculate them to first order with respect to a straight centered crack for an arbitrary trajectory in a strip of finite width. We are then able  to determine the instability threshold and the wavelength of the oscillations using  only the PLS  without introducing any additional criterion. Results agree very well  with the numerical simulations using  the phase field model. And comparing them   with  the results of~\citet{adda-95}, we find a small correction for the instability threshold, and a significant one for the wavelength. Also, unlike \citet{bouchbinder-03}, we argue that the calculation of the SIFs can be combined with the standard crack propagation criteria (that is with the Irwin criterion and the PLS) to obtain an evolution equation for the crack tip. Our main conclusion is that the PLS provides a good description of crack paths without any additional criteria. Recently, a similar approach for the thermal crack problem has been proposed and solved numerically~\citep{bahr-08} which is in agreement with both our theoretical findings and with experimental results of~\citet{ronsin-98}.

The paper is organized as follows. We first provide a description of the thermal crack problem. We then describe the phase field approach for this problem, and compare its results with those of a theoretical analysis we perform. An important result of the theoretical analysis is an evolution equation for the crack path which allows quantitative predictions. We conclude by discussing the relevance of this work to path prediction of cracks in general.

\section{The thermal crack problem}\label{problem}

\begin{figure}
\begin{center}
\includegraphics[height=3.5cm]{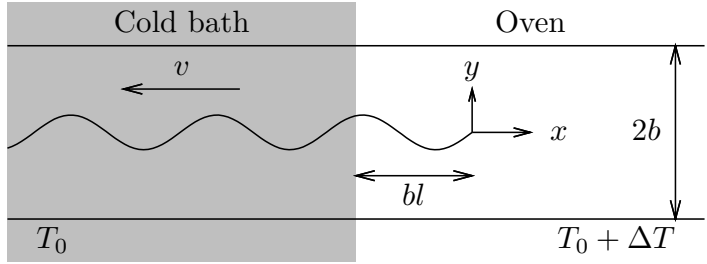}
\caption{The experimental setup. A glass plate of width $2b$ is moved slowly at a constant velocity $v$ from an oven maintained at temperature $T_0+\Delta T$ to a cold bath of fixed temperature $T_0$. The crack tip is steady in the laboratory frame (moving in the plate frame) and its position with respect to the cold bath ($bl$) is controlled by the temperature field.
\label{diagram}}
\end{center}
\end{figure}

We first introduce the thermal crack problem and set the basic definitions and notations. Here, we consider the idealized problem of an infinitely long strip containing a semi-infinite crack. The coordinate system is chosen so that the axis of symmetry of the strip corresponds to $y=0$ and the tip of the crack lies at $x=0$ (see Fig.~\ref{diagram}). We assume that the temperature field  is uniform across the thickness of the strip, and that the strip   is under plane stress conditions, so that the problem is actually  two-dimensional. We will also use dimensionless variables, measuring the lengths in units of the half-width of the strip $b$, the temperature in units of $\Delta T$, the strains in units of  $\alpha_T\Delta T/b$, and the stresses in units of $E\alpha_T\Delta T$, where $E$ is the Young's modulus and $\alpha_T$ the coefficient of thermal expansion. As will be shown, the behavior of the system is mainly governed by two dimensionless parameters~: the Peclet number $P=bv/D$, where $D$ is the thermal diffusion coefficient and the dimensionless material toughness $\hat K_\mathrm{Ic}=K_\mathrm{Ic}/(E\alpha_T\Delta T\sqrt b)$.

Provided the velocity of the plate is not too low, the temperature field, which is assumed not to be affected by the presence of the crack, can be described  by the advection-diffusion equation $ \partial_{xx}T+P\partial_x T = 0$, with the boundary conditions: $T(x)=0$ in the cold bath and $T(x)=1$ for $x\rightarrow\infty$~\citep{marder-94}. The solution of this equation is
\begin{equation}
T(x)=\left(1-e^{-P(x+l)}\right)\Theta(x+l)\;,
\label{Temperature}
\end{equation}
where $x=-l$ corresponds to the location of the surface of the cold bath (see Fig.~\ref{diagram}) and $\Theta(x)$ is the Heaviside function. Recall that the temperature field as given by Eq.~(\ref{Temperature}) is an approximation of the experimental one. For a quantitative comparison with experiments, the real temperature field should be used \citep{bahr-08}.

We consider the problem of a crack propagating through the heated strip within the framework of linear elastic fracture mechanics. Under plane stress conditions, the two-dimensional stress tensor $\overline{\overline{\sigma}}$ is related to the two-dimensional strain tensor $\overline{\overline{\epsilon}}$ by
\begin{equation}
\sigma_{ij}=\frac{1}{1-\nu^2}\left[(1-\nu)\epsilon_{ij}+\nu \epsilon_{kk}\delta_{ij}-(1+\nu)T(x)\delta_{ij}\right]\;,
\end{equation}
where $\nu$ is the Poisson ratio and the strain tensor $\overline{\overline{\epsilon}}$ is related to the displacement field $\vec{u}$ by
\begin{equation}
\epsilon_{ij}=\frac{1}{2}\left[\frac{\partial u_i}{\partial x_j}+\frac{\partial u_j}{\partial x_i}\right]\;.
\end{equation}
The strip is free from external traction, thus within any approach one should always satisfy the boundary conditions
\begin{equation}
\sigma_{yy}(x,\pm 1)=\sigma_{xy}(x,\pm 1)=0\;.
\label{eq:bc1}
\end{equation}

\section{The phase field model}
\label{modele_et_numerique}

The classical theory of crack propagation, where the crack behavior is
determined by the singularities of the stress field at the crack tip leads to
difficult numerical issues when considering the movement and interaction of
many cracks or to track crack motion in three dimensional systems where even
the dynamics of crack fronts is not well understood. In this context, the phase
field approach to crack
propagation~\citep{igorphasefield,Karmafrac,sethnaphasefield,Jagla} is very
useful as it allows to study problems in arbitrary geometries and go beyond
simple crack paths. It has succeeded in reproducing qualitative behavior of
cracks such as branching \citep{Karmabranching} and oscillations under biaxial
strain \citep{henrylevine2004}. More recently, it has been shown that the Griffith criterion and the PLS are embedded in the phase field model of crack propagation \citep{hakim08}.

The general idea of phase field modeling is to introduce an additional field (the phase field) that describes the state of the system. The main advantage of this approach is that one does not need to treat the interface explicitly since it is defined implicitly as an isosurface of the phase field. This advantage becomes clear when studying the evolution of complex shapes since the algorithmic cost of using an additional field is much smaller  than the cost of dealing explicitly with complex moving surfaces. In the case of fracture, this \textit{phase field} $\phi$ indicates whether the material is intact ($\phi=1$) or broken ($\phi=0$). In the usual sharp interface representation, a crack surface is an infinitely thin boundary between a region where the material is intact and an empty region that can not sustain any stress. In contrast, the equations of the phase field model are such that $\phi$ varies continuously in space, and a crack surface is represented by a region of finite thickness. The thickness can be chosen freely and does not affect numerical results as long as it is much smaller than the characteristic length scale of the studied phenomenon. It can be shown that through an appropriate coupling between the phase field $\phi$ and the elastic fields, the model can have the desired properties~: no stresses are transmitted across a crack (in the limit where the system size is much larger than the width of the diffuse interface) and one can associate a finite surface energy with the interface, so that the fracture energy is well-defined.

Here, we use this model to describe a quasistatic crack propagating due to a thermal gradient. We begin by presenting the model, including the adjustments needed to incorporate thermoelastic effects. The dimensionless elastic energy is written as~\citep{adda-95}
\begin{equation}
E_{el}=\int\int dxdy\,\,
g(\phi)\left[\frac{1}{4(1-\nu)}\left(\epsilon_{kk}- 2T(x)\right)^2+\frac{1}{2(1+\nu)}
\left(\epsilon_{ij}-\frac{\delta_{ij}}{2}\epsilon_{kk}\right)^2\right]\;,
\end{equation}
where the function $g(\phi)=\phi^3(4-3\phi)$ describes the coupling between the
phase field and the elastic fields. This function is such that in regions where
the material is broken ($\phi=0$), the contribution to the elastic energy is
zero, while in regions where the material is intact, the contribution to the
elastic energy recovers the one prescribed by linear elasticity. In~\citep{Karmafrac}, it is shown that the particular choice of $g$ does not affect the results as long as $g(0)=g'(0)=g'(1)=0$ and $\lim_{\phi=0}g(\phi)\sim\phi^\alpha$, with $\alpha>2$. The additional term $T(x)$ corresponds to the thermal expansion. This effect is assumed to be the same in the intact and in the partially broken parts of the material.

Since we are considering a quasi-static crack growth, in which the elastic body is at mechanical equilibrium at all times, the elastic energy should be an extremum with respect to the displacement field $\vec{u}(x,y)$. That is
\begin{equation}
\frac{\delta E_{el}}{\delta u_i}\label{el_eq}=0\;.
\end{equation}
We now need to specify the evolution of the phase field. This is done by associating with the phase field a dimensionless \textit{free energy} given by~\citep{henrylevine2004}
\begin{equation}
E_\phi=\int\int dxdy\left[\frac{D_\phi}{2}(\nabla\phi)^2+V(\phi)+
g(\phi)\left(\mathcal{E}_\phi- \mathcal{E}_c\right)\right]\;,
\end{equation}
where $\mathcal{E}_\phi$ is defined by
\begin{equation}
\mathcal{E}_\phi=\left\{
\begin{array}{lll}
\frac{1}{4(1-\nu)}\left(\epsilon_{kk}- 2T(x)\right)^2+\frac{1}{2(1+\nu)}
\left(\epsilon_{ij}-\frac{\delta_{ij}}{2}\epsilon_{kk}\right)^2
&\mbox{ if }& (\epsilon_{ii}-2T(x))>0\\
\frac{1}{2(1+\nu)}
\left(\epsilon_{ij}-\frac{\delta_{ij}}{2}\epsilon_{kk}\right)^2
&\mbox{ if }& (\epsilon_{ii}-2T(x))<0
\end{array}\;,
\right.
\end{equation}
where $V(\phi)=1.5 h (1-\phi^2)\phi^2$ is a double well potential,  with an energy barrier of height $\propto h$, that ensures
that the preferred states of the homogeneous system are either $\phi=1$ (intact)
or  $\phi=0$ (completely broken).
$\mathcal{E}_\phi$ coincides with the standard elastic energy density when the
material is under tension and includes only the contribution of shear when the
material is under compression. This choice avoids the propagation of a crack
under compression~\citep{henrylevine2004}. Taking into account the coupling
through the function $g(\phi)$, when
$\mathcal{E}_\phi$  is larger than a threshold value $\mathcal{E}_c$, the broken
phase is favored while when $\mathcal{E}_\phi<\mathcal{E}_c$, the intact phase
is favored. The evolution equation for $\phi$ is relaxational with a kinetic bias that ensures that the material will not be driven from a broken state to an intact state~:
\begin{equation}
\tau \dot{\phi}=\min\left(-\frac{\delta E_\phi}{\delta \phi},0\right)\;,
\label{eq_phidot}
\end{equation}
where $\tau$ is a constant. The irreversibility of Eq.~(\ref{eq_phidot}) means
that a crack cannot heal (which would happen in our set up  at the crack
tail) and ensures that the total elastic energy stored in the
material cannot increase when $\phi$ varies.

The elastic equation (\ref{el_eq}) is solved on a rectangular domain $[-\frac{L}{2},\frac{L}{2}]\times[-1,1]$ with $L\gg1$ using the Gauss-Seidel over-relaxation method at each time step~\citep{numerics}. In addition to the boundary conditions~(\ref{eq:bc1}), one has to set additional boundary conditions at $x=\pm L/2$. Here we have chosen
\begin{eqnarray}
\epsilon_{ij}(-L/2,y)&=&0 \,,\\
\sigma_{xy}(L/2,y)&=&\sigma_{xx}(L/2,y)=0\;.
\end{eqnarray}
It is clear that the final solution is insensitive to these boundary conditions
as long as $L\gg 1/P$. The phase field equation (\ref{eq_phidot}) is solved
using an Euler forward scheme~\citep{numerics}. The values of the grid spacing $dx$ and of the diffusion coefficient $D_\phi$ are chosen so
that the diffuse interface occupies few grid points and is much smaller than the characteristic wavelength of the oscillating crack, while the characteristic time $\tau$ is chosen so that the phase-field relaxes fast enough to its
equilibrium (or more precisely, multiplying $\tau$ by a factor $2$ does not
affect quantitatively the results.). Numerical checks were performed to ensure
that neither changes in the convergence criterion of the Gauss-Seidel iterative
method nor changes in the grid resolution affect the results. The
size of the domain $[L] \times [2]$ was typically $1500 \times 300$ grid points,
and numerical checks showed that doubling $L$ did not affect the results.
Finally, the fracture energy $\Gamma=
K_\mathrm{Ic}^2/(2\mu)$ was computed using the expression~\citep{Karmafrac}
\begin{equation}
 \Gamma=\int_0^1 d\phi \sqrt{2 D_\phi(\mathcal{E}_c-(\mathcal{E}_c g(\phi)-V(\phi))}\,.
\end{equation}
leading to the dimensionless material toughness $\hat K_{Ic}=\sqrt{2\mu\Gamma}/(E\alpha_T \Delta T \sqrt{b})$.

Before turning to the description of numerical results, we find it
necessary to summarize the different parameters used here and to
give some rationale for their choice. In addition to the physical parameters associated to the material and to the temperature field, the pertinent phase field parameters are $\tau$, $D_\phi$, $h$ and $\mathcal{E}_c$. In dimensionless units the values of $\tau=1$, $D_\phi=4.4\times10^{-3}$, $h=1$ and $\mathcal{E}_c=1$ have been chosen. Notice that the three latter parameters determine both the fracture energy and  the interface width  of the phase field. Since we are considering the dimensionless material toughness  $\hat K_{Ic}$, the only effect we are concerned with, is the dependance of the width of the diffuse interface (the broken surface) on $D_\phi$. More precisely, in the spirit of the phase field approach, the value of $D_\phi$, for a given value of  $\hat K_{Ic}$ should not affect the behaviour of the crack. Within the present simulations, we have checked that using a value of $D_\phi$ corresponding to an interface which is twice thinner does not affect
quantitatively the final results.

\begin{figure}
\begin{center}
\includegraphics[height=4.cm]{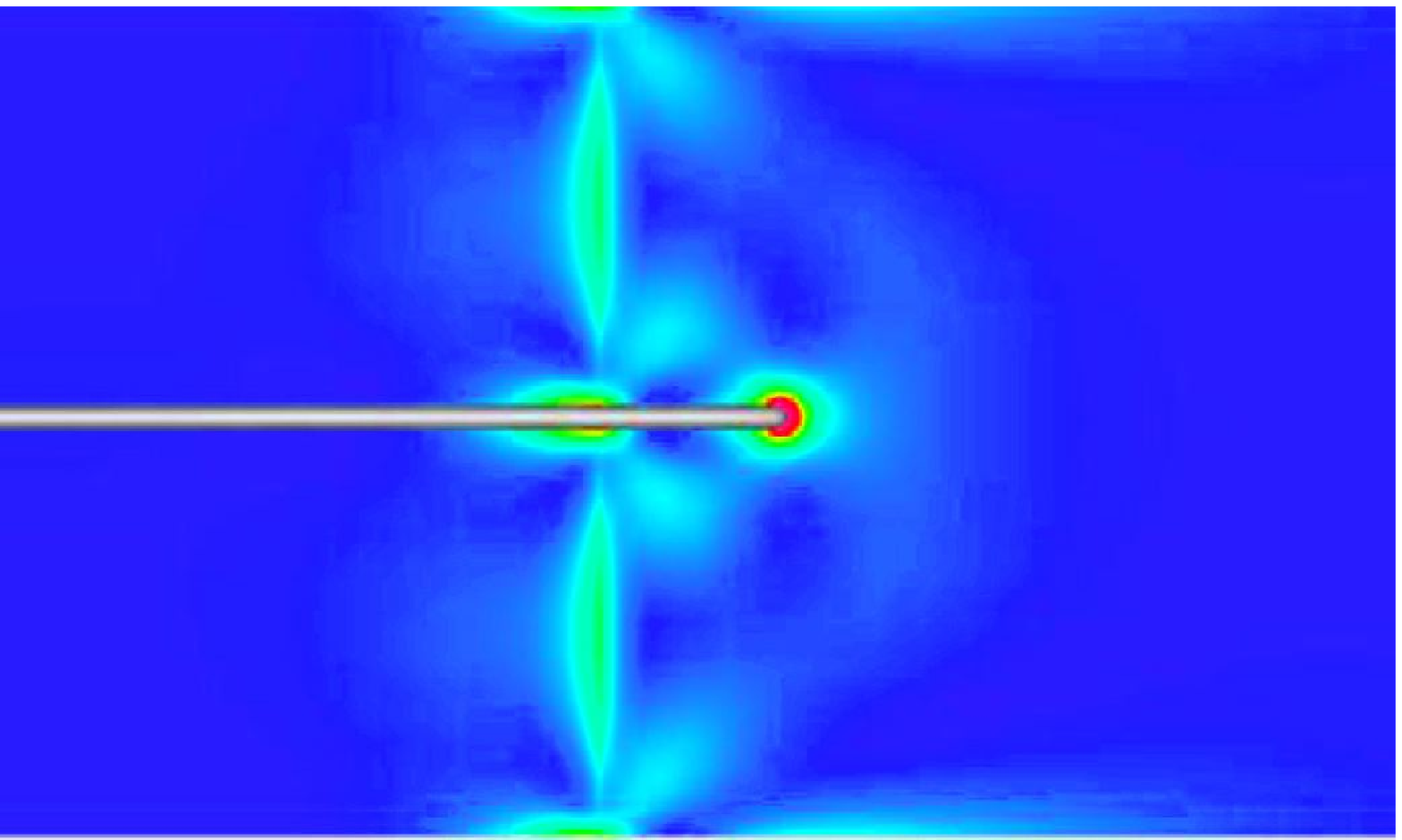}
\includegraphics[height =4.cm]{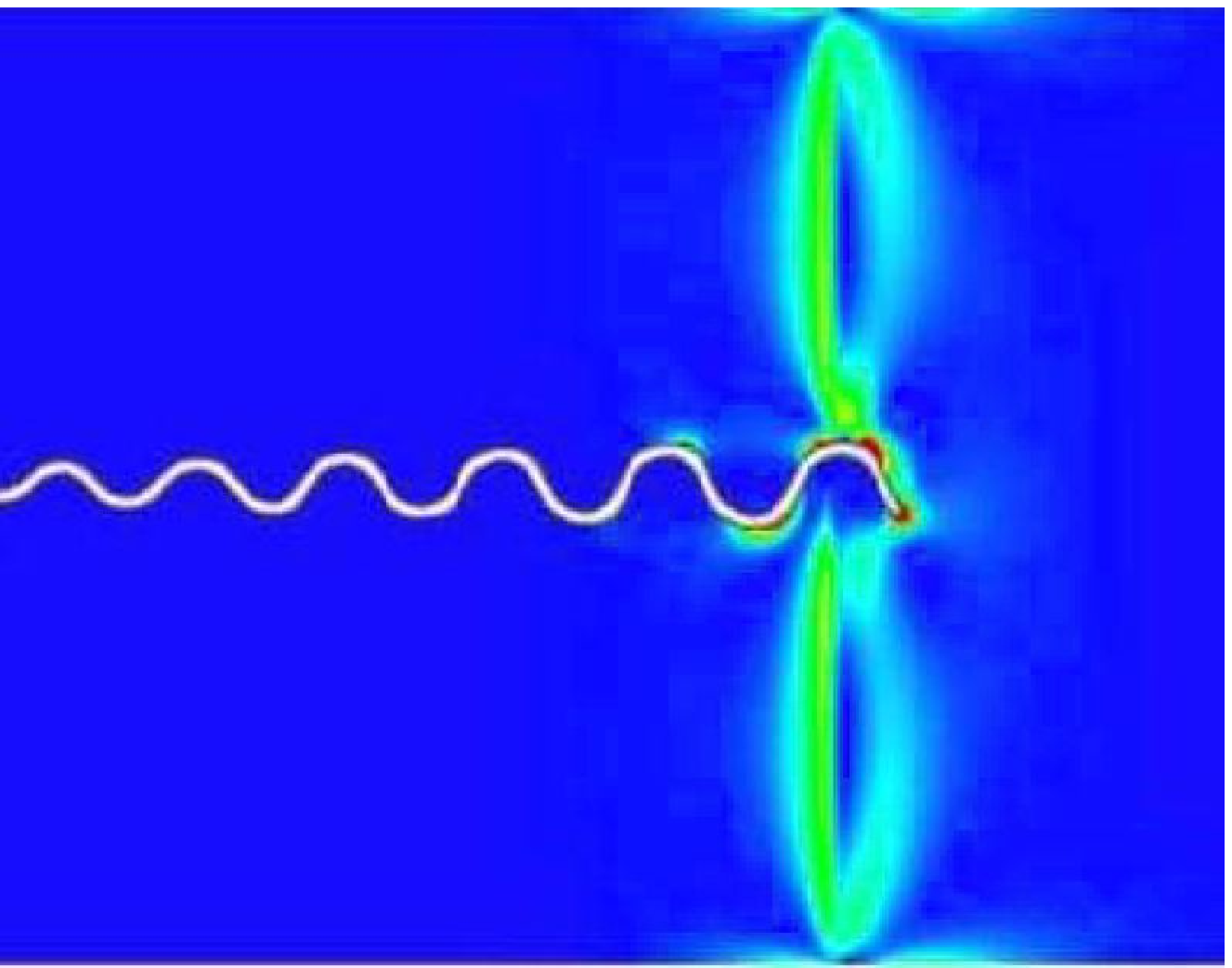}
\caption{Typical crack paths obtained using the phase field model.  The white region corresponds to the broken surface and the color code corresponds to the density of elastic energy in the material (red: high elastic energy density, blue: zero elastic energy). \textbf{(a)} A straight centered propagating crack. \textbf{(b)} A wavy crack above the instability threshold. This picture shows that the crack width is much smaller than the wavelength of the observed oscillations.
\label{fig_trans}}
\end{center}
\end{figure}

We now summarize the results of the simulations. By tuning the control parameters $P$ and $\hat K_\mathrm{Ic}$, we could observe the different regimes observed in experiments~: no crack propagation, propagation of a straight crack at the center of the strip or a crack following a wavy path (see Fig.~\ref{fig_trans}). However, the main purpose of the present simulations was to determine the nature of the bifurcation from straight to oscillatory cracks, since no direct experimental results are available. Namely in experiments, the amplitude of oscillations varies almost linearly with the control parameter, but since experimental results do not allow to explore the behavior of the crack close to the threshold it was impossible to characterize the crack path in the close vicinity of the transition point. Furthermore, it should be emphasized that apart from attempts describing this transition by a Landau-Ginzburg expansion \citep{bahr-95,bouchbinder-03}, the available theoretical work has been mostly limited to linear stability analysis. Using the phase field model and without resorting to any assumption, we were able to compute the steady-state paths and to determine the nature of the bifurcation.

\begin{figure}
\begin{center}
\includegraphics[height=2.72cm]{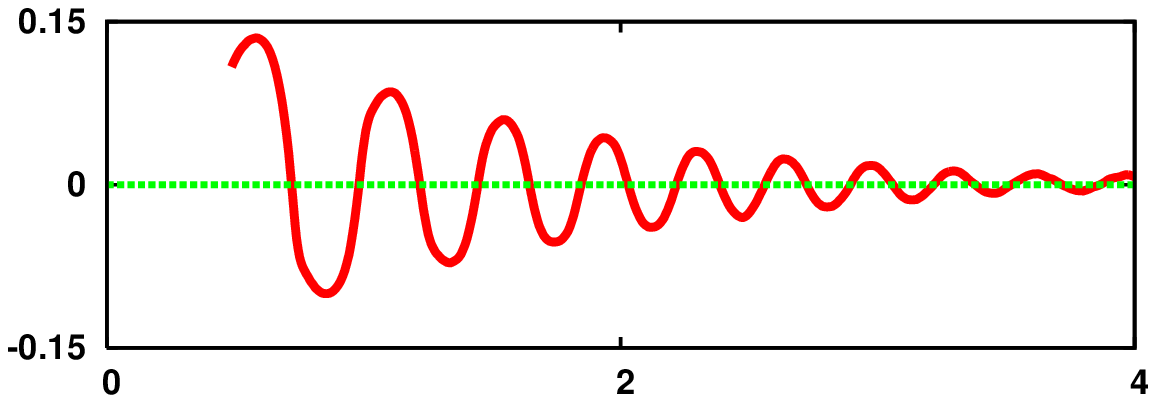}
\includegraphics[height=2.72cm]{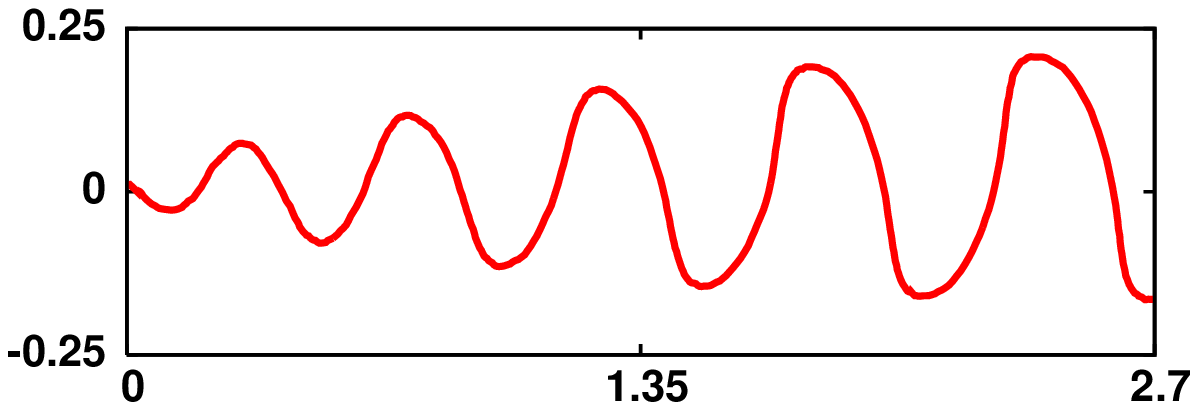}
\caption{\textbf{(a)}:  Crack tip trajectory below the threshold of linear instability for a crack which was initially off center. $P=7.9$ and $1/\hat K_\mathrm{Ic}=7.85$. \textbf{(b)}: Crack tip trajectory for an initially slightly off center crack above the threshold of linear instability. $P=7.9$ and $1/\hat K_\mathrm{Ic}= 8.43$.
\label{fig_paths}}
\end{center}
\end{figure}

\begin{figure}
\begin{center}
\includegraphics[height=5.55cm]{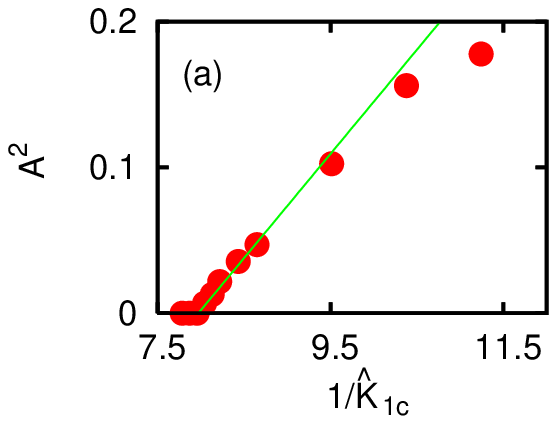}
\includegraphics[height=5.55cm]{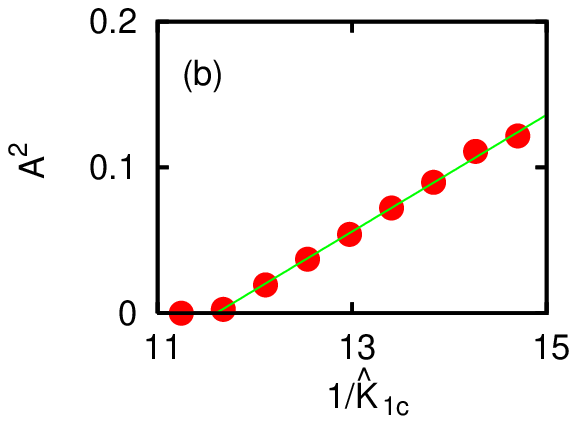}
\caption{ \textbf{(a)} Square of the amplitude $A$ of the oscillations for $P=7.89 $ as function of $1/\hat K_\mathrm{Ic}$, which is related to the amplitude of the temperature gradient. \textbf{(b)}: same as \textbf{(a)} with $P=3.95$. The linear behavior of $A^2$ in the neighborhood of the transition shows that the transition from straight to oscillatory propagation is supercritical (continuous).
\label{fig_amplis}}
\end{center}
\end{figure}

First, in Fig.~\ref{fig_paths} we present some typical crack paths close to the
threshold for the sake of comparison with the analytical results presented below
and also in order to clarify the nature of the bifurcation. The first path was
obtained with an initial straight crack which was off center and below the
threshold. One can see damped oscillations as the crack path returns to its
equilibrium state: a straight crack. The second path was obtained with an
initial crack that was {\it slightly} off center (shifted by $2$ grid points
away from the center) and {\it slightly} above the threshold. One can see
oscillations of the path that  are  amplified as the crack advances. These
oscillations eventually saturate and the crack path becomes oscillating with a
finite constant  amplitude as observed in experiments. Using the model we were
able to compute the amplitude of the oscillations when a steady state was
reached. The simulations showed that the steady state is
independant of initial conditions, such as the initial position of the crack, and
is a single valued function of the control parameters. Typical results are shown in Fig.~\ref{fig_amplis} where the square of
the amplitude is plotted as a function of $1/\hat{K}_{1c}$ for a given value of
$P$. One can see that above a threshold value of  $1/\hat{K}_{1c}$ the amplitude
of oscillations is no longer zero and scales as  the square root of the
deviation from the threshold.  The same behavior holds when using the value of
$P$ as a control parameter. The amplitude curves together  with the damped
oscillations below the threshold and the amplified oscillations above  the
threshold indicate that the bifurcation is supercritical or continuous.

\begin{figure}
\begin{center}
\includegraphics[height=3cm]{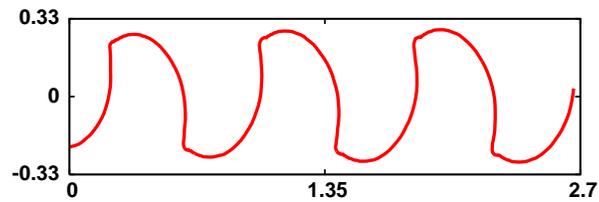}
\caption{Crack tip trajectory for a crack deep in the nonlinear regime, that is far above the threshold of the instability. The values of the control parameters are $P=7.9$ and $1/\hat K_\mathrm{Ic}=11.6$. Note the striking resemblance to crack paths appearing in~\citep{ghatak,audoly,sendova}.
\label{fig_paths2}}
\end{center}
\end{figure}

For values of the control parameters well above the threshold, we were not able to recover complex paths similar to those observed in experiments~\citep{yuse-93}. This may be due to the fact that well above the threshold the quasistatic approximation is no longer valid and dynamic effects through the temperature field as given by Eq.~\ref{Temperature} become relevant. It is nonetheless interesting to note that the observed crescent-like path in Fig.~\ref{fig_paths2} exhibits a striking similitude with paths obtained when cutting a thin elastic sheet with a moving thick object~\citep{ghatak,audoly} and with crack paths obtained during drying of silicate sol-gel films fabricated using spin-coating techniques~\citep{sendova}.

\begin{figure}
\begin{center}
\includegraphics[height=6.8cm]{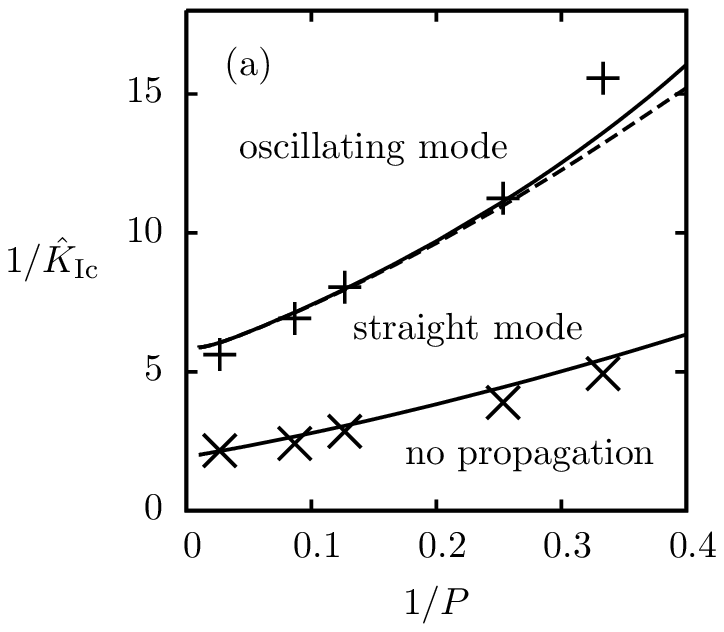}
\includegraphics[height=7cm]{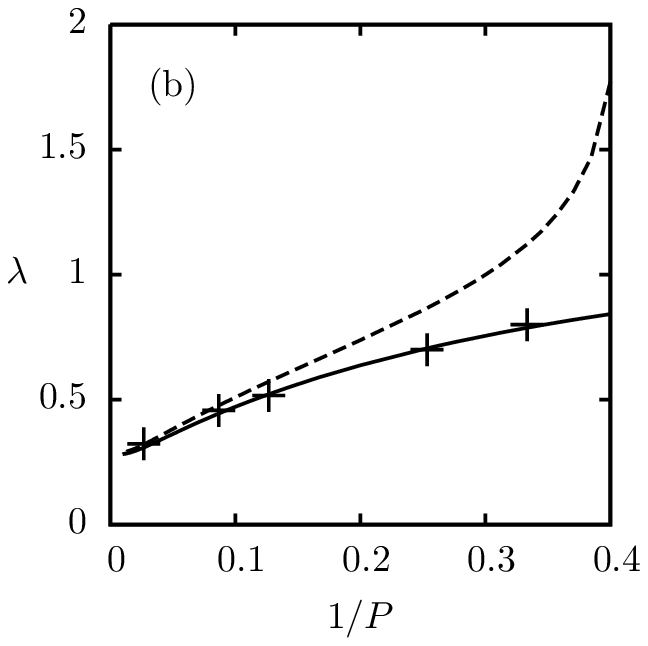}
\caption{Main theoretical and numerical results. (a) The phase diagram of the different states of the system as a function of the two control parameters $P$ and $\hat K_\mathrm{Ic}$. (b) The wavelength $\lambda$ as a function of $P$ at the transition from straight to wavy crack propagation. In both figures, the symbols $+$ and $\times$ correspond to the numerical results obtained from the phase field model, the solid lines correspond to the analytical calculation presented in Sec.~\ref{sec:theory} and the dashed lines correspond to the results reported in \citep{adda-95}.
\label{seuils}}
\end{center}
\end{figure}

The phase diagram in Fig.~\ref{seuils}(a) shows the threshold for the propagation of a centered straight crack and for the transition to a wavy crack propagation in the $1/P$--$1/\hat{K}_{1c}$ phase space.  It can be seen that the theoretical predictions presented in Sec.~\ref{sec:theory} are in quantitative agreement with  numerical results of the phase field model without any adjustable parameters. The same agreement is reached when considering the dimensionless wavelength at the threshold of the transition from straight to wavy crack path (see Fig.\ref{seuils}(b)). For this latter quantity, one should also note the good agreement with experimental values for small values of $1/P$ (linear behavior with  $\lambda\simeq0.28+2.0/P$) \citep{RonsinPRL}. Moreover, the phase field computation and our theoretical approach lead to the same deviation from the linear behavior for $1/P>0.2$, where the discrepancy between the present results and those reported in \citep{adda-95} become significant. This feature has also been observed in experiments~\citep{Ronsinthese}.

\section{Theoretical analysis}
\label{sec:theory}

We now turn to a theoretical study of the transition from straight to oscillatory crack propagation. As in previous theoretical treatments \citep{marder-94,adda-95,bouchbinder-03}, the analysis is based on the calculation of the stress intensity factors (SIFs). However, here, they are computed to first order for an arbitrary trajectory without using the infinite plate approximation and without the restriction that the crack tip lies at the center of the sample. Within the framework of linear elastic fracture mechanics, the propagation of a crack is governed by the singular behavior of the stress field in the vicinity of its tip, as defined in Eq.~(\ref{SIF}). The propagation laws we use are the Irwin criterion defined in Eq.~(\ref{Irwin}) and the principle of local symmetry (PLS) defined in Eq.~(\ref{eq:PLS0}). To calculate the SIFs, one must solve the equilibrium equations
\begin{equation}
\frac{\partial\sigma_{ij}}{\partial x_j}=0\;,\qquad
\nabla^2\sigma_{ii}=-\nabla^2T(x)\;,
\label{eq:bulk}
\end{equation}
with the boundary conditions (\ref{eq:bc1}) at the sides of the plate and the additional boundary condition at  the crack surface
\begin{equation}
\sigma_{ij}n_j=0\;,
\label{eq:bc2}
\end{equation}
where $\overrightarrow n$ is the normal to the crack faces.

To investigate the transition from a straight to an oscillating regime, we consider a small deviation from a straight centered crack,
\begin{equation}
y(x)=A h(x)\;,
\end{equation}
where $h(x)$ is defined for $x\le 0$ (see Fig.~\ref{diagram}) and $|A|\ll 1$. We expand the displacement and stress fields to first order as
\begin{eqnarray}
u_i&=&u_i^{(0)}+A u_i^{(1)}+O\left(A^2\right)\;,\label{eq:expandu} \\
\sigma_{ij}&=&\sigma_{ij}^{(0)}+A\sigma_{ij}^{(1)}+O\left(A^2\right)\;. \label{eq:expands}
\end{eqnarray}
The SIF $K_\mathrm{I}$ (resp. $K_\mathrm{II}$) is an even (resp. odd) function of $A$. Therefore their expansions have the form
\begin{eqnarray}
K_\mathrm{I}&=&K_\mathrm{I}^{(0)}+O(A^2)\;,
\label{eq:perturb1}\\
K_\mathrm{II}&=&A K_\mathrm{II}^{(1)}+O(A^3)\;.
\label{eq:perturb2}
\end{eqnarray}
Note that in particular, $K_\mathrm{II}=0$ for a straight centered crack. The different terms of the above expansion can be determined by solving the equilibrium equations using repeatedly the Wiener-Hopf method. The details of these calculations are presented in Appendix~A. Here we simply summarize  the calculation of $K_\mathrm{II}^{(1)}$, which is novel. $K_\mathrm{II}^{(1)}$ can be written as
\begin{equation}
\label{kappadef}
K_\mathrm{II}^{(1)}[h(x)]=\kappa h(0)+K_\mathrm{II}^{(1)}[h(x)-h(0)]\;,
\label{eq:K2h}
\end{equation}
which amounts to decomposing the general problem into that of a straight off-center crack and that of an oscillating crack with a centered tip. The latter problem was solved by~\citet{adda-95}, yielding
\begin{equation}
K_\mathrm{II}^{(1)}[h(x)-h(0)]=\frac{K_\mathrm{I}^{(0)}}{2}h'(0)
+\int_{-\infty}^{0}dx\frac{\partial}{\partial x}\left[\sigma_{xx}^{(0)}(x,0)(h(x)-h(0))\right]p^+(-x)\;,
\end{equation}
where $p^+(x)$ is a weight function that does not depend on the physical parameters. We have generalized the results of~\citet{adda-95} with the calculation of the contribution $\kappa h(0)$ to obtain
\begin{equation}
K_\mathrm{II}^{(1)}[h(x)]=\left[cK_\mathrm{I}^{(0)}-\frac{\partial}{\partial l}K_\mathrm{I}^{(0)}\right]h(0)
+\frac{K_\mathrm{I}^{(0)}}{2}h'(0)
+\int_{-\infty}^{0}dx\frac{\partial}{\partial x}\left[\sigma_{xx}^{(0)}(x,0)h(x)\right]p^+(-x)\;,
\label{eq:K2h2}
\end{equation}
where $c\simeq1.272$ is a numerical constant (see Appendix A for details).

\begin{figure}
\begin{center}
\includegraphics[height=7cm]{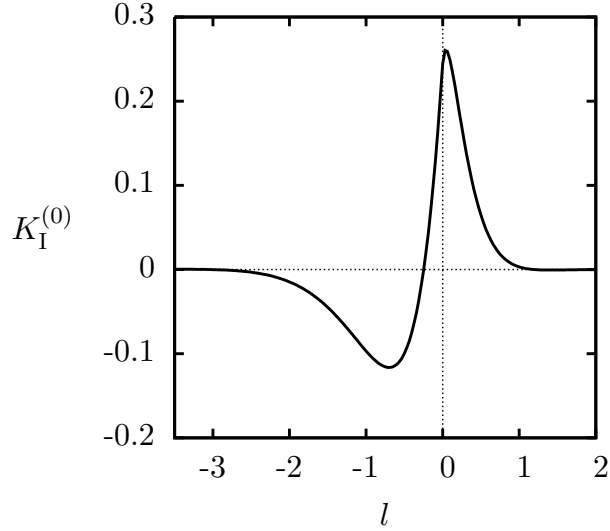}
\caption{$K^{(0)}_\mathrm{I}$ as a function of $l$, for $P=5$. Recall that $l$ is the distance of the crack tip from the boundary of the cold bath. The threshold of incipient growth of a centered straight crack is given by $\hat K_{\mathrm{Ic}}=K^{(0)}_\mathrm{I}(l_c)$, where $l_c$ is the location of the maximum of $K^{(0)}_\mathrm{I}(l)$.
\label{KIplot}}
\end{center}
\end{figure}

Having calculated the SIFs to first order, we can now examine the particular form taken by the crack propagation criteria for a small perturbation. The Irwin criterion reads $K_\mathrm{I}^{(0)}=\hat
K_\mathrm{Ic}$. The transition to a propagating straight crack is thus governed by the existence of $l$ such that $K_\mathrm{I}^{(0)}(l)=\hat K_\mathrm{Ic}$. This occurs when $\hat K_{\mathrm{Ic}}=K^{(0)}_\mathrm{I}(l_c)$, where $l_c$ is the location of the maximum of $K^{(0)}_\mathrm{I}(l)$  (see Fig.~\ref{KIplot}). This condition determines the transition curve in Fig.~\ref{fig_trans} from no crack propagation to a propagation of a centered straight crack. Also, above the propagation threshold Eq.~(\ref{eq:perturb1}) shows that  to first order in $A$, the distance between the cold front and the crack tip, which is determined by $K_I\left[ h(x) \right]$, does not depend on the crack path $y(x)=A h(x)$. This simplifies the problem, as we can parameterize the ``time" evolution of the crack path by the position of its tip in a coordinate system attached to the plate. If $h(x)$ now describes the crack path in such a coordinate system, the value of $K_\mathrm{II}^{(1)}$ when the tip is at the point $x$ is simply given by
\begin{equation}
\label{KIIx}
K_\mathrm{II}^{(1)}(x)=\left[cK_\mathrm{I}^{(0)}-\frac{\partial}{\partial l} K_\mathrm{I}^{(0)}\right]h(x)
+\frac{K_\mathrm{I}^{(0)}}{2}h'(x)
+\int_{-\infty}^0du\frac{\partial}{\partial u}\left[\sigma_{xx}^{(0)}(u,0)h(x+u)\right]p^+(-u)\;.
\end{equation}
Therefore, the PLS which for a small perturbation reads $K_\mathrm{II}^{(1)}(x)=0$, yields
\begin{equation}
0=\left[cK_\mathrm{I}^{(0)}-\frac{\partial}{\partial l} K_\mathrm{I}^{(0)}\right]h(x)
+\frac{K_\mathrm{I}^{(0)}}{2}h'(x)
+\int_{-\infty}^{0}du\frac{\partial}{\partial u}\left[\sigma_{xx}^{(0)}(u,0)h(x+u)\right]p^+(-u)\;.
\label{eq:PLS}
\end{equation}
Eq.~(\ref{eq:PLS}) is a linear equation for a crack that deviates slightly from the straight centered path.

To determine the stability of the straight centered crack propagation, we must examine the evolution of a small perturbation, which amounts to solving the following problem~: given a crack path $h(x)$ defined for $x\le 0$, which does not necessarily satisfy $K_\mathrm{II}^{(1)}(x)=0$ for $x<0$, find a continuation of $h$ such that $K_\mathrm{II}^{(1)}(x)=0$ for $x>0$. Now, since $\sigma_{xx}^{(0)}(x,0)$ decreases exponentially as $x\to-\infty$, one can expect the long-term behavior of the crack to be the same as that of a crack path satisfying $K_\mathrm{II}^{(1)}(x)=0$ for all $x$, and this is indeed observed in the simulations (see Fig.~\ref{fig_paths}(b) in the previous section, and the discussion below). Therefore, examining crack paths that satisfy $K_{II}^{(1)}=0$ from the beginning is sufficient to determine the stability threshold, and we will use such paths for that purpose.

\begin{figure}
\begin{center}
{\includegraphics[height=7cm]{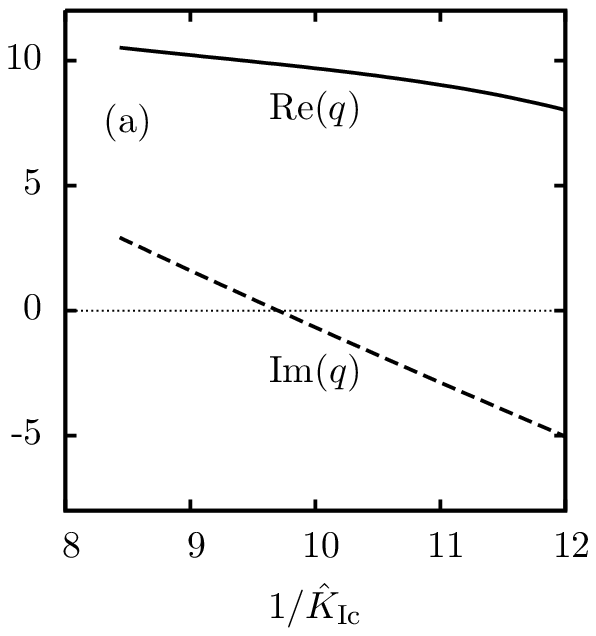}}
{\includegraphics[height=7cm]{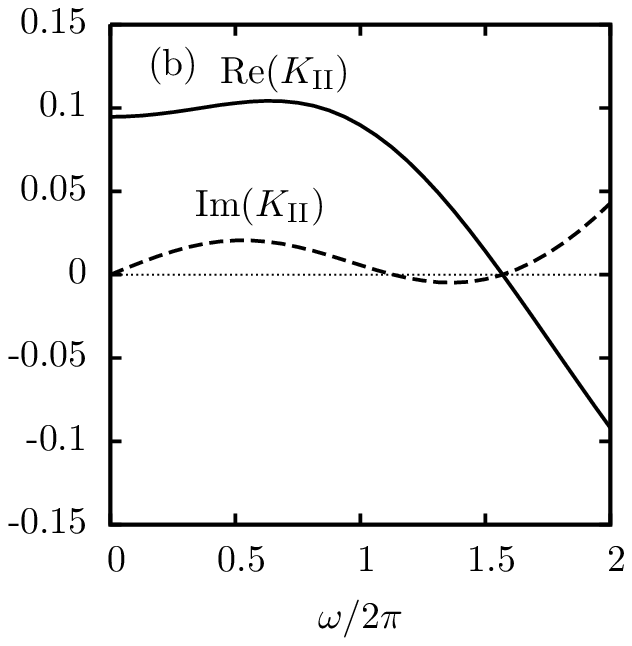}}
\caption{(a) Solutions of $ K_\mathrm{II}(q)=0$ versus $1/\hat K_\mathrm{Ic}$ for $P=5$. The solution branch disappears when $\Im(q)$ exceeds a certain value, as the integral appearing in the calculation of $ K_\mathrm{II}(q)$ ceases to converge. (b) The real and imaginary part of $K_\mathrm{II}(\omega)$, with $\omega$ real, at the instability threshold for $P=5$ ($\hat K_\mathrm{Ic}\simeq0.103)$.
\label{modes2}}
\end{center}
\end{figure}

We look for solutions of the form
\begin{equation}
h(x)=\Re\left[\delta \,\exp(iqx)\right]\;,
\label{eq:hq}
\end{equation}
where $\Re$ stands for the real part, $\delta$ and $q$ are in general complex numbers. Then using Eq.~(\ref{KIIx}), $K_\mathrm{II}^{(1)}(x)$ takes the form
\begin{equation}
K_\mathrm{II}^{(1)}(x)=\Re\left[\delta \, K_\mathrm{II}(q)\,\exp(iqx)\right]\; ,
\end{equation}
where
\begin{equation}
 K_\mathrm{II}(q)=\left[cK_\mathrm{I}^{(0)}-\frac{\partial}{\partial l} K_\mathrm{I}^{(0)}\right]
+iq\frac{K_\mathrm{I}^{(0)}}{2}
+\int_{-\infty}^{0}du \frac{\partial}{\partial u}\left[\sigma_{xx}^{(0)}(u,0)e^{iqu}\right]p^+(-u)\;.
\label{eq:dispersion}
\end{equation}
A solution $K_\mathrm{II}^{(1)}(x)=0$ is obtained if $ K_\mathrm{II}(q)=0$, which is the dispersion relation of the problem. Since $q$ and $ K_\mathrm{II}(q)$ are in general complex numbers, we can expect the equation $ K_\mathrm{II}(q)=0$ to admit a discrete
set of solutions for a given set of the dimensionless parameters $P$ and $\hat K_\mathrm{Ic}$. By examining the behavior of $ K_\mathrm{II}(q)$, we find that the stability is determined by
solutions corresponding to a pair of conjugate oscillating modes, which are shown on Fig.~\ref{modes2}(a) for a given value of $P$. When $\Im(q)>0$, the perturbation (\ref{eq:hq}) is damped which implies that the straight and centered crack propagation is stable. When $\Im(q)<0$, the perturbation is amplified, leading to a wavy crack propagation regime. Thus the instability threshold corresponds to $\Im(q)=0$, and is denoted by $q\equiv q_{c}$. Using this result, another way to determine the instability threshold is illustrated in Fig.~\ref{modes2}(b). It consists in supposing from the beginning that $q\equiv\omega$ is a real number\footnote{we use $\omega$ instead of $q$ in order to emphasize the fact that it is a real number} and in looking for the specific value of $\omega=q_c$ for which $\Im[ K_\mathrm{II}(\omega)]=0$ and $\Re[ K_\mathrm{II}(\omega)]=0$ simultaneously.

By repeating the above computation for different values of $P$ and $\hat K_\mathrm{Ic}$, we obtained in Fig.~\ref{seuils}(a) the phase diagram and in Fig.~\ref{seuils}(b) the wavelength of the oscillations at the threshold, which is given by $\lambda=2\pi/q_{c}$. These figures show that a good quantitative agreement is reached with the numerical results of the phase field model for both the wavy instability threshold and the wavelength at the transition.

Our results for the instability threshold differ from those of~\citet{adda-95}, in which the threshold was determined by the existence of a real solution $q_{c}$ such that $K_\mathrm{II}^{(1)}[\sin \omega x]=\frac{dK_\mathrm{II}^{(1)}}{d\omega}[\sin \omega x]=0$ for $x=0$ and the stability criterion was based on the sign of $K_\mathrm{II}^{(1)}[\sin \omega x]$. In general, such a solution has no reason to satisfy $K_\mathrm{II}^{(1)}(x)=0$ for $x\ne 0$. Indeed, the wavelengths predicted by these two criteria (i.e., the present one and the one of \cite{adda-95}) differ significantly, especially for small values of $P$  (see Fig.~\ref{seuils}(b)). The thresholds predicted also differ, but less notably (see Fig.~\ref{seuils}(a)). As can be seen in Fig.~\ref{modes2}(b), the deviations between the estimates of $\hat K_\mathrm{Ic}$ and $\lambda$  obtained by the two methods are related to the vertical and horizontal distances between the minimum of $\Im(K_\mathrm{II})$ and its intersection with $\Re(K_\mathrm{II})$, and the vertical distance  scales as the square of the horizontal one.

\section{Conclusions}

In this paper we address the problem of crack path prediction for
quasi-static crack propagation. We used the framework of the thermal
crack experiment since it is a well controlled experiment, and the best known to us. The
reason is that stresses are induced internally by an imposed
thermal gradient that can be easily tuned by external parameters
leading to a steady state propagation.

Our main result is that by using only the Principle of Local
Symmetry and the Griffith criterion one can reproduce the phase
diagram of the problem, namely predict the existence of the two
instabilities (no-crack to straight crack and straight to
oscillatory propagation). Furthermore, we show that these criteria
are sufficient to determine the instability threshold and the
wavelength of the oscillations without introducing additional
hypotheses. Interestingly, our theoretical results agree very well with the numerical simulations of the phase field model. This strengthens the theoretical approach based on the classical LEFM theory combined with the two propagation criteria, namely PLS and Griffith criterion (but mostly PLS). This work joins a recent effort to explain a variety of instabilities of cracks, namely the branching instability \citep{katzav07b} and the out-of-the-plane roughening instability \citep{katzav07a} using these general criteria (i.e., LEFM + Griffith criterion + PLS), 

Another achievement of this work is the fact that the phase field model allows to clarify the nature of the transition between straight and oscillatory cracks, which is convincingly shown to be supercritical or continuous. This is achieved thanks to the high resolution of the phase field approach near the transition point and
the relative ease at which in-silico experiments can be performed under different conditions. This also suggests that the phase field model is more adapted than the classical methods for crack path tracking, especially when considering complex moving shapes, in view of its relatively low computational cost. It is therefore a good
candidate to describe complicated crack patterns such as spirals or crescents as the ones observed in \citep{sendova}, branching \citep{katzav07b,henry2008} and interacting cracks \citep{Jagla}. A real challenge for this approach would be to reproduce propagation of cracks in 3D, which can help to infer their laws of motion from such well controlled numerical experiments.

\appendix

\section{The straight off-center crack}

In the following, we present briefly the method of resolution for a straight off-center crack in a thermal gradient. The method of resolution follows the same steps as in~\citep{adda-95}. Consider a straight crack which is slightly above/below the center of the strip, namely its location is given by $y(x)=Ah(0)\equiv \delta $. Due to the absence of symmetry with respect to the center of the strip $y=0$, let us split the displacement, strain and stress fields in the strip as follows
\begin{equation}
f(x,y)=\frac{1}{2}\left(f^+(x,y) \Theta(y-\delta)+f^-(x,y)\Theta(\delta-y)\right)\;,
\end{equation}
where $\Theta$ is the heaviside function and $f$ stands for any component of the elastic fields. Let us also define
\begin{equation}
[f](x)=\frac{1}{2}\left(f^+(x,\delta)-f^-(x,\delta)\right)\;,
\end{equation}
Note that all the elastic fields are continuous in the unbroken regions. However, along the crack surfaces  the displacements might be discontinuous while $\sigma_{yy}$ and $\sigma_{xy}$ are continuous there. Thus, for this specific problem, the boundary conditions (\ref{eq:bc1},\ref{eq:bc2}) can be rewritten as
\begin{eqnarray}
\sigma^{\pm}_{yy}(x,\pm 1)&=&\sigma^{\pm}_{xy}(x,\pm 1)=0\;,\
\label{eq:bc3}\\
\left[\sigma_{yy}\right](x)&=&\left[\sigma_{xy}\right](x)=0 \;,
\label{eq:bc4}\\
\left[u_{x}\right] (x)&=&\left[u_{y}\right](x)=0 \qquad \textrm{for}\quad x>0 \; ,
\label{eq:bc5}\\
\sigma^{\pm}_{yy}(x,\delta)&=&\sigma^{\pm}_{xy}(x,\delta)=0 \qquad \textrm{for}\quad x<0 \; ,
\label{eq:bc6}
\end{eqnarray}

Now, we Fourier transform along the $x$ direction all the elastic fields and the temperature field and plug them into the equilibrium equations (\ref{eq:bulk}) and into the boundary conditions (\ref{eq:bc3},\ref{eq:bc4}). After some algebraic manipulations, this yields relations of the type
\begin{equation}
\left( {\begin{array}{*{20}c}
   {\left[ {\hat u_x } \right](k)}  \\
   {\left[ {\hat u_y } \right](k)}  \\
\end{array}} \right) = \left( {\begin{array}{*{20}c}
   {C_{xx}(k) } & {C_{xy}(k) } & {D_x(k) } \\
   {C_{yx}(k) } & {C_{yy} (k)} & {D_y(k) }  \\
\end{array}} \right)\left( {\begin{array}{*{20}c}
   {\hat \sigma _{xy}(k,\delta) }  \\
   {\hat \sigma _{yy}(k,\delta) }  \\
   \hat T(k) \\
\end{array}} \right) \, ,
\label{eq:matrix}
\end{equation}
Performing the first-order expansions in $\delta$, as given by Eqs.~(\ref{eq:expandu})-(\ref{eq:expands}):
\begin{eqnarray}
\hat u_i(k,\delta) &=& \hat u_i^{(0)}(k,0)+\delta \hat u_i^{(1)}(k,0) \;,\label{eq:expanduk} \\
\hat\sigma_{ij}(k,\delta)&=& \hat\sigma_{ij}^{(0)} (k,0)+\delta \hat \sigma_{ij}^{(1)}(k,0)    \;, \label{eq:expandsk}
\end{eqnarray}
and inverting (\ref{eq:matrix}) gives the final result
\begin{eqnarray}
\label{order0}
\hat\sigma_{yy}^{(0)}(k,0)&=&-F(k)\left[\hat u_y^{(0)}\right](k,0)+D(k)\hat T(k)\;,\\
\label{sigmaxx}
\hat\sigma_{xx}^{(0)}(k,0)&=&H(k)\hat\sigma_{yy}^{(0)}(k,0)+S(k)\hat T(k)\;,\\
\label{order1}
\hat\sigma_{xy}^{(1)}(k,0)&=&
    -P(k)\left(\left[\hat u_x^{(1)}\right](k,0)-ik\left[\hat u_y^{(0)}\right](k,0)\right)+ik\hat\sigma_{xx}^{(0)}(k,0)\;,
\end{eqnarray}
where
\begin{equation}
\begin{array}{ll}
F(k)=\displaystyle{k\frac{\sinh^2 k-k^2}{\sinh 2k+2k}}\,,&
D(k)=\displaystyle{2\frac{(1-\cosh k)(\sinh k-k)}{\sinh 2k+2k}}\;,\\
H(k)=\displaystyle{\frac{k^2+\sinh^2 k}{\sinh^2 k-k^2}}\;,&
S(k)=\displaystyle{\frac{\sinh k-k}{\sinh k+k}}\;,\\
P(k)=\displaystyle{k\frac{\sinh ^2 k-k^2}{\sinh 2k-2k}}\;.
\end{array}
\end{equation}

The first step of resolution which involves only the zeroth-order expansion, i.e. the case of a straight centered crack, is the calculation of $K_\mathrm{I}^{(0)}$. Using the boundary conditions~(\ref{eq:bc5},\ref{eq:bc6}) to leading order and applying the Wiener-Hopf method to equation (\ref{order0}) yields~\citep{adda-95}
\begin{equation}
K_\mathrm{I}^{(0)}=\int_{-\infty}^{+\infty}D(k)\hat T(k)F^+(k)\frac{dk}{2\pi}\;,
\end{equation}
where
\begin{equation}
F(k)=\frac{F^-(k)}{F^+(k)}\;,
\end{equation}
and $F^+(k)$ ($F^-(k)$) is analytic in the upper (lower) half-plane, respectively.

To complete the calculation of the SIFs, we return to the case of a straight off-center crack and determine the coefficient $\kappa$ appearing in Eq.~(\ref{kappadef}). Using the boundary conditions~(\ref{eq:bc5},\ref{eq:bc6}) to first order in $\delta$ and applying again the Wiener-Hopf method to Eq.~(\ref{order1}) yields
\begin{eqnarray}
\left[\hat u_x^{(1)}\right](k)&=&ik\left[\hat u_y^{(0)}\right](k)
    -\frac{1}{P^-(k)}\int_{-\infty}^{+\infty}
    \frac{ik'P^+(k')\hat\sigma_{xx}^{(0)-}(k',0)}{k'-k+i\epsilon}\frac{dk'}{2i\pi}
    +\frac{\alpha}{P^-(k)}\nonumber\\
&=&-\frac{k}{F^-(k)}\int_{-\infty}^{+\infty}
    \frac{D(k')\hat T(k')F^+(k')}{k'-k+i\epsilon}\frac{dk'}{2\pi}\nonumber\\
&&-\frac{1}{P^-(k)}\int_{-\infty}^{+\infty}
    \frac{ik'P^+(k')\hat\sigma_{xx}^{(0)-}(k',0)}{k'-k+i\epsilon}\frac{dk'}{2i\pi}
    +\frac{\alpha}{P^-(k)}\;,
\end{eqnarray}
where $P(k)=P^-(k)/P^+(k)$ with  $P^+(k)$ ($P^-(k)$) analytic in the upper (lower) half-plane,
and $\epsilon$ is a vanishingly small positive number. notice that the solution of $\left[\hat u_y^{(0)}\right](k)$ obtained from the zeroth order calculations has been used, and that $\sigma_{xx}^{(0)}(x,0)$ can be obtained from Eq.~(\ref{sigmaxx}). For both $F(k)$ and $P(k)$ the factorization into $F^\pm(k)$ and $P^\pm(k)$ can be achieved using a Pad\'e decomposition, as previously explained in~\citep{bouchbinder-03,katzav07b}. The unknown constant $\alpha$ is introduced because the decomposition involved in the Wiener-Hopf method is not unique. The value of $\alpha$ can be determined by noting that the $\left[\hat u_x^{(1)}\right](k)$ behaves as $O(k^{-3/2})$ as $k\to\infty$. Canceling out the $k^{-1/2}$ term on the r.h.s. yields
\begin{equation}
\alpha=-\int_{-\infty}^{+\infty}D(k)\hat T(k)F^+(k)\frac{dk}{2\pi}\;,
\end{equation}
and thus one has
\begin{eqnarray}
\left[\hat u_x^{(1)}\right](k)&=&\left(\frac{1}{F^-(k)}-\frac{1}{P^-(k)}\right)
    \int_{-\infty}^{+\infty}D(k')\hat T(k')F^+(k')\frac{dk'}{2\pi}\nonumber\\
    &&+\frac{1}{F^-(k)}\int_{-\infty}^{+\infty}
    \frac{-ik'D(k')\hat T(k')F^+(k')}{k'-k+i\epsilon}\frac{dk'}{2i\pi}\;.
\end{eqnarray}
Finally by looking at asymptotic behavior of $\left[\hat u_x^{(1)}\right](k)$ as $k\to\infty$ one has
\begin{equation}
\label{kappa}
\kappa=cK_\mathrm{I}^{(0)}-\frac{\partial}{\partial l} K_\mathrm{I}^{(0)}
-\int_{-\infty}^{0}\frac{\partial}{\partial x} \sigma_{xx}^{(0)}(x,0)p^+(-x)dx\;,
\end{equation}
where $p^+(x)$ is the inverse Fourier transform of $P^+(k)$ and $c$ is a numerical constant given by
\begin{equation}
c=\lim_{k\to\infty}\frac{(ik)^{\frac{3}{2}}}{\sqrt{2}}\left[\frac{1}{F^-(k)}-\frac{1}{P^-(k)}\right]\simeq1.272\;.
\end{equation}
Using this value of $\kappa$ in Eq.~(\ref{eq:K2h}) yields Eq.~(\ref{eq:K2h2}).

\end{document}